\documentclass[twocolumn]{oldjpsj2} 
\usepackage[dvips]{color}

\title{Mott Transition in the Hubbard Model\\
on Checkerboard Lattice}

\author{Takuya \textsc{Yoshioka}$^{1}$\thanks{E-mail address: yoshioka@tp.ap.eng.osaka-u.ac.jp}, Akihisa \textsc{Koga}$^{2}$ and Norio \textsc{Kawakami}$^{2}$}

\inst{$^{1}$Department of Applied Physics, Osaka University, Suita, Osaka 565-0871, Japan \\
$^{2}$Department of Physics, Kyoto University, Kyoto 606-8502, Japan}

\abst{We investigate the bandwidth-controlled Mott transition in the Hubbard model on the checkerboard lattice at half filling using the path-integral renormalization group (PIRG) method. It is demonstrated that the system undergoes a first-order phase transition to the plaquette-singlet insulating phase at a finite Hubbard interaction. This conclusion is drawn via a detailed analysis of the spin and charge correlations around the phase transition point by means of the PIRG method aided with a new iteration scheme introduced in this paper. 
}

\kword{Mott transition, Hubbard model, checkerboard lattice, geometrical frustration, path integral renormalization group method (PIRG)}

\begin{document}
\maketitle
\section{Introduction}
\begin{figure}[htb]
  \begin{center} 
\includegraphics[width=0.45\textwidth]{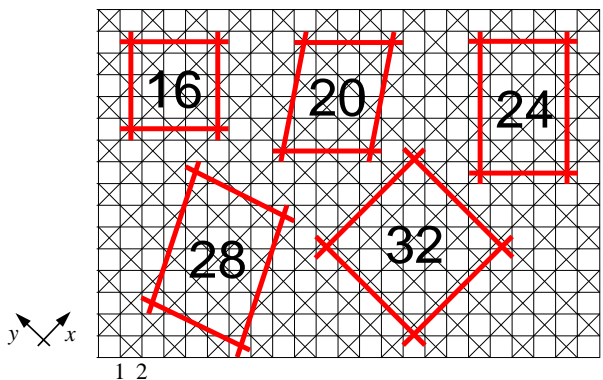}
\caption{(Color online) Checkerboard lattice. 
Boxes indicate the clusters used in this paper and the number 
labels the size of them.
}
\label{fig1}
  \end{center}
\end{figure}

Strongly correlated electron systems with geometrical frustration have attracted much interest recently. A well-known example is the frustrated pyrochlore lattice, which is given by a three-dimensional corner-sharing network of tetrahedra. This family includes the transition-metal oxides $\rm LiV_2O_4$ \cite{exp3} and ${\rm Tl_2Ru_2O_7},$\cite{exp1,exp2} where heavy fermion behavior and the Mott transition without magnetic ordering were observed at low temperatures. In these compounds, electron correlations on the frustrated lattice may be a source of intriguing properties at low temperatures. These experimental findings have stimulated the intensive studies of the Hubbard model on the pyrochlore lattice and its two-dimensional (2D) analog called the checkerboard lattice (Fig. \ref{fig1})\cite{Isoda,Fujimoto2,Fujimoto,tJ,LiV1,LiV2,LiV3,LiV4,LiV5,LiV6,LiV7,LiV8,LiV9}.
We will focus on the checkerboard lattice model in this paper. 

\begin{figure}[htb]
  \begin{center}
\includegraphics[width=0.45\textwidth]{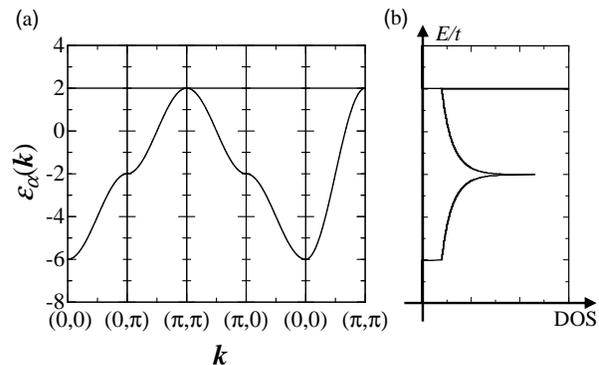}
\caption{(a) The band structure along the symmetry lines in the Brillouin zone (B.z.) and (b) the density of states in the 
noninteracting case on the checkerboard lattice (eq. (2)).
}
\label{fig2}
  \end{center}
\end{figure}

The pyrochlore- and checkerboard-lattice models with nearest-neighbor hopping have some peculiar properties at half filling; the Fermi level just touches a flat band (Fig. \ref{fig2}), which invalidates standard perturbation calculations \cite{Isoda,Fujimoto2,Fujimoto} and thus makes theoretical investigations extremely difficult. This in turn poses a challenging theoretical problem in the Mott physics in the frustrated systems. In the early work on the checkerboard lattice model, it was claimed that there may be a metal-insulator (MI) transition at infinitesimally small interaction, which is followed by the second insulator-insulator transition \cite{Fujimoto2,Fujimoto}. 
Later on, however, it was demonstrated that the paramagnetic metallic state is stabilized at least in the weak coupling region \cite{label5,label6}, although it remains still open how the paramagnetic metallic state competes with other states, and how it is connected to the plaquette valence-bond crystal state realized in the strong coupling limit \cite{Fouet}. 

In this paper, we investigate electron correlations in the Hubbard model on the checkerboard lattice at half filling at absolute zero by means of the path-integral renormalization group (PIRG) method. We show  that the system undergoes a single first-order phase transition to the plaquette-singlet Mott phase at a finite Hubbard interaction. To carry out the calculation, we propose a new iteration scheme for the PIRG algorithm, which substantially improves the numerical accuracy, and hence allows us to figure out the nature of the phase transition.

The paper is organized as follows. In \S2, we introduce the model Hamiltonian and briefly explain the PIRG method. The newly introduced iteration scheme is also mentioned in the section. We discuss the phase transition in the checkerboard Hubbard model in \S3. A brief summary is given in \S4.

\section{Model and Method}

We start with the Hubbard model on the checkerboard lattice (see Fig. \ref{fig1}),
\begin{equation}
\hat{\cal{H}}=-t\sum_{(im,jm'),\sigma}
\hat{c}^{\dagger}_{im\sigma}\hat{c}_{jm^{'}\sigma}
+U\sum_{i,m}\hat{n}_{im\uparrow}\hat{n}_{im\downarrow},
\end{equation}
where $\hat{c}_{im\sigma}$ ($\hat{c}^{\dagger}_{im\sigma}$) is an annihilation (creation) operator of electron in the $i$-th unit cell with spin $\sigma$ and sublattice index $m$ (=1,2), and $\hat{n}_{im\sigma}= \hat{c}^{\dagger}_{im\sigma}\hat{c}_{im\sigma}$. $U$ is the Coulomb interaction and $t$ is the transfer integral with the same couplings on vertical, horizontal and diagonal bonds of the checkerboard lattice. A unitary transformation of the kinetic term of the Hamiltonian $\hat{{\cal H}}_k$  gives the diagonalized form $\hat{{\cal H}}_k=\sum_{k,\alpha,\sigma}\varepsilon_{\alpha}({\mbox{\boldmath$k$}})\hat{a}^{\dagger}_{k\alpha\sigma}\hat{a}_{k\alpha\sigma}$,
with two eigenvalues for each ${\mbox{\boldmath$k$}}$,
\begin{equation}
\varepsilon_{\alpha}({\mbox{\boldmath$k$}})
=\left\{
\begin{array}{ll}
\displaystyle
2t\ &{\rm for}\ \alpha=1, \\
-2t(1+\cos k_x+\cos k_y)\ &{\rm for}\ \alpha=2,
\end{array}
\right.
\end{equation}
where $\alpha$ represents the band index. We assume $t>0$, hereafter. The dispersion relations and the corresponding density of states are depicted in Fig. \ref{fig2}. The topmost energy band is flat over the whole Brillouin zone (B.z.), while the lower one is dispersive with the characteristic band structure for the square lattice with nearest-neighbor hopping. It is known that the band structure originates from highly frustrated geometry of the checkerboard lattice \cite{graph}.

As discussed in Refs. \cite{Isoda, Fujimoto2}, the perturbation expansion in $U$ encounters divergence at third and higher orders, which is due to the presence of the flat band in the system. Therefore, powerful numerical techniques are necessary to discuss the phase transitions in the model. In the frustrated system,  the quantum Monte Carlo simulations usually suffer from the minus sign problems, and the exact diagonalization may not be efficient to discuss the ground state properties in the thermodynamic limit. Although one can deal with larger systems by means of the variational Monte Carlo simulations, the obtained results strongly depend on a trial wave function employed. In this way, it may not be easy to discuss the nature of the phase transitions in the system. To overcome the difficulty, we here make use of the PIRG method developed by Imada {\it et al.},\cite{PIRG0,PIRG1} which allows us to improve the ground state systematically. It has an advantage in treating frustrated electron systems with large clusters, in contrast to the other numerical methods.

The PIRG treatment is in principle independent of an initial state and an iterative method employed. However, careful choices of them are important to converge the PIRG calculations within available computational time.  To obtain an appropriate initial state, we make use of the unrestricted Hartree-Fock (UHF) approximation.\cite{QMCstart} In the approximation, the site- and spin-dependent mean fields $\langle \hat{n}_{im\sigma}\rangle$ are introduced and the interaction term $\hat{{\cal H}}_U$ in the Hamiltonian is then replaced by
\begin{eqnarray}
\hat{\cal{H}}_U^{\rm UHF}&=&U_{int}\sum_{i,m}[\langle\hat{n}_{im\uparrow}\rangle\hat{n}_{im\downarrow}+\hat{n}_{im\uparrow}\langle\hat{n}_{im\downarrow}\rangle \nonumber\\
&&-\langle\hat{n}_{im\uparrow}\rangle\langle\hat{n}_{im\downarrow}\rangle],
\end{eqnarray}
where the parameters $\langle \hat{n}_{im\sigma}\rangle$ are determined by the self-consistent equations 
$\langle \hat{n}_{im\sigma}\rangle=
\langle\phi_0|\hat{n}_{im\sigma}|\phi_0\rangle$. Here $|\phi_0\rangle$ is 
the ground state of the Hamiltonian 
$\hat{{\cal{H}}_k}+\hat{\cal{H}}_U^{\rm UHF}$.
We use the resulting wave function as an initial one for the PIRG simulations, and then take into account quantum fluctuations. Note that the interaction $U_{int}$ is not necessarily equal to the original $U$ when the initial wave function is determined. By performing the PIRG iteration, the approximate ground state is described by the Slater basis states as $|\psi\rangle=\sum_{\alpha=1}^Lc_\alpha|\phi_{\alpha}\rangle$, where $c_\alpha$ is an amplitude of $|\phi_{\alpha}\rangle$, and $L$ is the dimension of the truncated Hilbert space. 
Furthermore, by using an energy variance extrapolation scheme,
\cite{PIRG0,PIRG1,Sorella} 
we can deduce physical quantities such as the ground state energy and the double occupancy. Here the energy variance is defined by $[\langle\hat{{\cal H}^2}\rangle-\langle\hat{\cal H}\rangle^2]/\langle\hat{\cal H}\rangle^2$.

We now introduce a new iteration scheme which can substantially improve the convergence of PIRG algorithm for the frustrated systems. Note that the true ground state could be obtained in principle by acting the time evolution operator on the initial state as
\begin{eqnarray}
|\psi_g\rangle = e^{-\beta \hat{H}}|\phi_0\rangle,
\end{eqnarray}
with imaginary time $\beta\rightarrow\infty$. To approach the  ground state in the framework of the PIRG method, we divide the operator into those in a small imaginary time slice $\Delta\tau$. In terms of the Hubbard-Stratonovich transformations, the operator is explicitly given as
\begin{eqnarray}
e^{-\Delta\tau {\hat{\cal H}}}
&=&\sum_{\{s\}} e^{-\Delta\tau{\hat{\cal H}}_k/2}\ \hat{V}(\{s\})\ 
e^{-\Delta\tau {\hat{\cal H}}_k/2}\nonumber\\ 
&+&O((\Delta\tau)^3,\\
\hat{V}(\{s\})&=&\prod_{i=1}^N\frac{1}{2}e^{\alpha (s_i)\hat{n}_{i\uparrow}}e^{\alpha (-s_i)\hat{n}_{i\downarrow}},
\end{eqnarray}
where $\sum_{\{s\}}$ represents $\sum_{s_1=\pm1}\sum_{s_2=\pm1}\ldots
\sum_{s_N=\pm1}$, $\alpha(s)=2as-\Delta\tau U/2$ and 
$a=\tanh^{-1}\sqrt{\tanh\left(\Delta\tau U/4\right)}$. 

In the ordinary PIRG method, 
the Hilbert subspace with $L$ basis states is generated by 
the time evolution operators for the local interacting term 
$e^{-\Delta\tau U\hat{n}_{i\uparrow}\hat{n}_{i\downarrow}}$.
Since the operator is replaced by the potential terms with one auxiliary field as $e^{-\Delta\tau U\hat{n}_{i\uparrow}\hat{n}_{i\downarrow}}=1/2\sum_{s_i=\pm1}e^{\alpha (s_i)\hat{n}_{i\uparrow}}e^{\alpha (-s_i)\hat{n}_{i\downarrow}}$, 
it generates two basis states. Therefore, by acting the operators with different $i$ on a certain basis state a couple dozen times, we can easily produce $L$ states. However, it is difficult to numerically generate {\it independent} basis states in this procedure, because the effective dimension is often decreased. This drawback may result from the fact that two states generated by the local Hubbard-Stratonovich field are very similar, making it sometimes difficult to distinguish themselves within given numerical accuracy. This may cause the reduction in the effective dimension of the Hilbert subspace. To improve the situation, we here select $L$ basis states { \it randomly} within the $2^N$ states generated by the operator $e^{-\Delta \tau' \hat{H}}$ with the {\it full} Hamiltonian.
In this case, unfavorable history accumulated via successive actions of the projection operators does not appear and thereby we can effectively avoid the decrease in the dimension of Hilbert subspace. In particular, when $\Delta\tau'(\neq 0)$ is small, the above procedure improves the PIRG results substantially, and the introduction of random variables does not cause serious statistical errors.

After the initial basis states are generated in the above procedure, we have to modify these states to approach the true ground state. The ordinary PIRG iteration process, which is denoted as Iteration process B, is based on the local update. Therefore, the optimized ground state is sometimes trapped in a local minimum of the energy, and could be far from the true ground state. To overcome this problem, we introduce the following iteration process denoted as A, which enables us to modify the basis state globally, in spite of the local update.

The detail of both processes is summarized as follows.
\begin{itemize}
\item Iteration process A
\begin{itemize}
\item 
To approach the true ground state, we have to modify the basis states $\{|\phi_a\rangle\}$. We pick up one basis state $|\phi_a\rangle$, and compare it with its modified state 
$|\phi_a'\rangle=e^{-\Delta\tau{\hat{\cal H}}_k/2}\ 
\hat{V}(\{s\})\ e^{-\Delta\tau {\hat{\cal H}}_k/2}|\phi_a\rangle$.
If the energy for $|\phi_a'\rangle$ with appropriate sets of $\{s\}$ 
is lower than that for $|\phi_a\rangle$, we update the basis state. 
We perform similar updating procedures successively for the $L$ basis states.
\end{itemize}
\end{itemize}

\begin{itemize}
\item Iteration process B
\begin{itemize}
\item We perform the ordinary PIRG iteration procedure.\cite{PIRG0,PIRG1}
We first consider a new basis state by acting the following three operators
on a certain basis state $|\phi_a\rangle$ step by step: first
$e^{-\Delta\tau\hat{H}_{k}/2}$, and then 
$e^{-\Delta\tau U n_{i\uparrow} n_{i\downarrow}}$ for all $i$, and
finally $e^{-\Delta\tau\hat{H}_{k}/2}$. In each step, we pick up the lowest-energy state among the original state and the newly generated states. We perform  similar procedures for all the basis states.
\end{itemize}
\end{itemize}
In our PIRG calculation, we first perform the process A a couple dozen times
to modify the basis states globally. After that, the basis states are locally modified by the process B.
The combined scheme, which is referred to as 'Iteration A $\rightarrow$ B', 
allows us to approach the ground state efficiently in small iteration steps and thus obtain the physical quantities precisely.

Furthermore, to improve numerical accuracy, we also use the quantum-number projection scheme in the framework of the PIRG method (PIRG+QP).\cite{PIRG2} In this paper, we employ a projection operator to the total spin-singlet 
state, $\hat{P}_{S=0}$, which acts on the converged state obtained in the above procedure:
$|\psi_{S=0}\rangle=
\sum_{\alpha=1}^Lc_\alpha'\hat{P}_{S=0}|\phi_{\alpha}\rangle$.
Here $c_\alpha'$ is an amplitude reevaluated  from the generalized eigenvalue problem for a new set of the quantum-number projected basis states 
\{$\hat{P}_{S=0}|\phi_{\alpha}\rangle$\}.  The PIRG+QP method is particularly powerful to precisely determine the phase diagram of frustrated electron systems \cite{PIRG7}.

\begin{figure}[htb]
  \begin{center}
\includegraphics[width=0.45\textwidth]{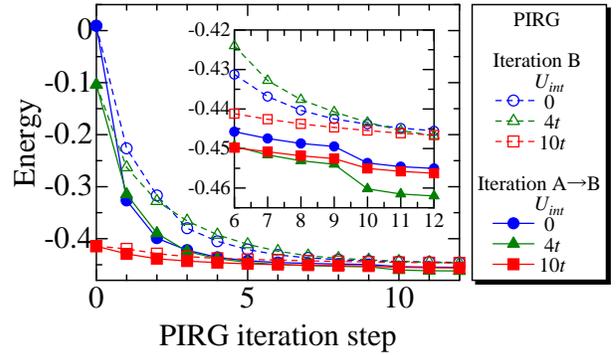}
\caption{(Color online) 
Energy as a function of the number of PIRG iteration at $L=500$ 
in the Hubbard model on checkerboard lattice of $N=16$ with $U=10t$. 
Circles, triangles and squares represent the results obtained 
from  different initial states.
For comparison, the results for several different iteration processes are shown. Note that the iteration mode is changed from A to B between the 
9th and 10th step in the  case of 'Iteration A$\rightarrow$B'.
}
\label{fig3}
\end{center}
\end{figure}
Here, we demonstrate how effective our method introduced here is. In Fig. \ref{fig3}, we show the energy as a function of the number of iteration for the small cluster of $N=16$ with $U=10t$, which is obtained by 
the PIRG method with $L=500$, $\Delta\tau=0.5$, and $\Delta\tau'=\Delta\tau\times 10^{-3}$. It is found that the energy converges in a dozen steps of iteration in spite of the fact that a large number of basis states are treated. This is in contrast to the results of the ordinary PIRG method 
where a few hundred steps of iteration are needed for the convergence. 
Note that the converged ground state is not necessarily optimized by the initial state obtained by the UHF approximation with $U=U_{int}$. For example, in the present case $U=10t$, the optimized ground state is deduced more effectively starting from the UHF approximation with $U_{int}=4t$. For comparison, we also show the results obtained only by the Iteration B process. It is found that the energy is always slightly higher than the energy obtained by 'Iteration A $\rightarrow$ B'. Although the energy difference seems small, this optimization process of the ground state plays a crucial role in deducing the physical quantities precisely.
\begin{figure}[htb]
  \begin{center}
\includegraphics[width=0.45\textwidth]{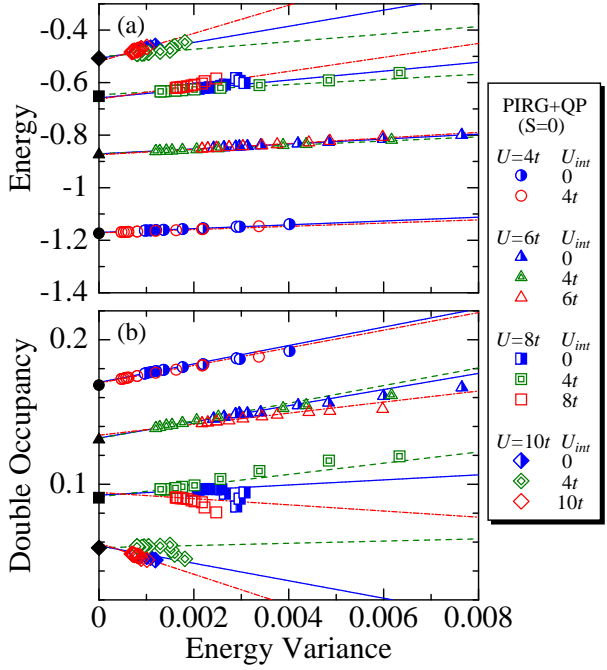}
\caption{(Color online) 
(a) Energy and (b) double occupancy 
per site as a function of the energy variance with various initial states
 in the Hubbard model on checkerboard lattice $(N=16)$. 
For comparison, exact results are shown as solid symbols.}
\label{fig4}
\end{center}
\end{figure}
We show the results for the cluster of $N=16$ with $U=10t$ obtained by the PIRG+QP method with a few choices of $U_{int}$. In fact, it is found that when the energy variance is small, the data in each $U_{int}$ are well fitted to a straight line, as shown in Fig. $\ref{fig4}$. We thus  obtain the ground state energy and the double occupancy precisely, and confirm that the PIRG+QP results are indeed in good agreement with those of the exact diagonalization.

In the following, we carry out PIRG+QP calculations for the half-filled systems with $N=16,\ 20,\ 24,\ 28,$ and $32$ sites in periodic boundary conditions (see Fig. \ref{fig1}). We note that in the non-interacting case, the cluster with $N=20$ has a closed shell structure while the others an open shell structure. Therefore, in the latter, the highest energy level in the lower band and the energy level of the flat band are degenerate. This means that the Fermi point appears only at ${\mbox{\boldmath$k$}}={\mbox{\boldmath$k$}}_F[=(\pi,\pi)]$ in the normal metallic state. Although this "quasi-Fermi" point  artificially appears under the periodic boundary condition, it might be a good probe to observe the MI transition.
In the PIRG method, we use $\Delta\tau\times U/t=0.5$ for both iteration processes and $\Delta\tau'=\Delta\tau\times 10^{-3}$ for generating initial states, and repeat the iteration until the energy converges under the truncated Hilbert space. We keep the Slater basis states up to $L=500$. Note that numerical errors for the finite cluster (thermodynamic limit), which will be shown in the next section, mostly arise from the energy variance (system size) extrapolation scheme.

\section{Results}
\begin{figure}[htb]
  \begin{center}
\includegraphics[width=0.45\textwidth]{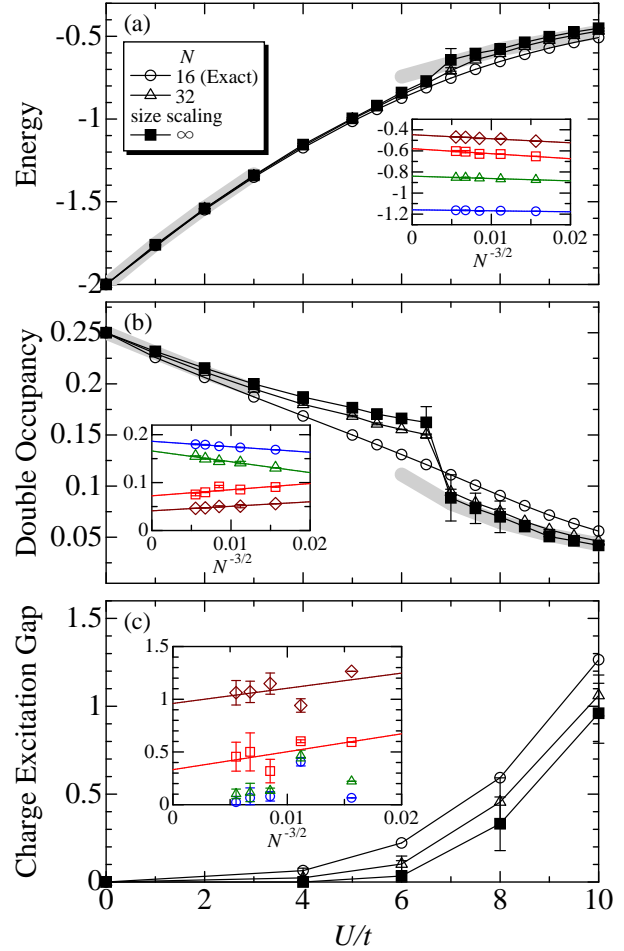}
\caption{(Color online) 
(a) The ground state energy $E_g/N$, (b) the double occupancy $(\partial E_g/\partial U)/N$ and (c) the charge excitation gap $\Delta_c$ as a function of $U/t$ on $N=16$ (circle) and $32$ (triangle) lattice and in the thermodynamic limit (square). Thick gray lines in (a) and (b) are the results obtained by the weak and strong coupling techniques (see text). 
Just $N=16$ sample is obtained by exact diagonalization.
$N=\infty$ results are obtained from finite size scaling, 
some of which are shown in the inset for $U/t=4$ (circle), $6$ (triangle), $8$ (square) and $10$ (diamond). We note that $N=20$ system has closed shell structure at half filling so $\Delta_c=0.382$ even $U=0$.}
\label{fig5}
\end{center}
\end{figure}

In this section, we present the numerical results for the half-filled Hubbard model on the checkerboard lattice. First, we compute the ground state energy, the double occupancy and the charge excitation gap. The results are shown in Fig. \ref{fig5}.
We also perform a finite-size scaling to obtain the results in the thermodynamic limit, as shown in the inset, where a scaling form of $N^{-3/2}$ is used for the physical quantities. In the noninteracting case, the dispersive band $(\alpha=2)$ is fully occupied, where $E_g/N=-2t$ and $(\partial E_g/\partial U)/N=0.25$. The introduction of the Coulomb interaction monotonically decreases the double occupancy, as expected for the paramagnetic metallic state which persists up to $U/t\sim 6$\cite{label5,label6}. Further increase in the interaction yields a cusp singularity in the ground state energy and a jump singularity in the double occupancy, as shown in Figs. \ref{fig5} (a) and (b). This suggests the existence of the first-order phase transition. The transition point is estimated as $U_c/t=6.75\pm0.25$. To discuss the nature of the transition, we also calculate the charge excitation gap $\Delta_c$, which is defined by a difference between two chemical potentials,
\begin{equation}
\Delta_c=\frac{\mu_+-\mu_-}{2},
\end{equation}
where $\mu_+=\left[E_g(M_\uparrow+1,M_\downarrow+1)-E_g(M_\uparrow,M_\downarrow)\right]/2$, $\mu_-=\left[E_g(M_\uparrow,M_\downarrow)-E_g(M_\uparrow-1,M_\downarrow-1)\right]/2$ 
and $E_g(M_\uparrow,M_\downarrow)$ is the ground state energy of the system,
where $M_\sigma$ is the number of electrons with spin $\sigma$.
We find in Fig. \ref{fig5} (c) that the charge gap is zero for $U<U_c$ while it is finite for $U>U_c$, i.e. the system is driven to an insulating state. It is, however, difficult to quantitatively estimate the charge gap in the thermodynamic limit from the finite-size scaling analysis within the present numerical accuracy, as seen in the inset of Fig. \ref{fig5} (c). 

\begin{figure}[htb]
  \begin{center}
\includegraphics[width=0.45\textwidth]{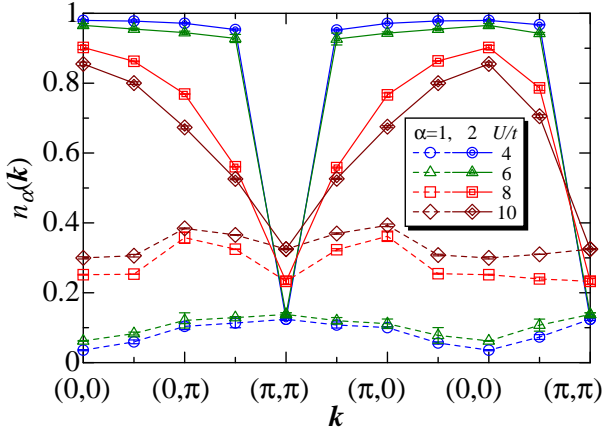}
\caption{(Color online) 
The momentum distribution function 
$n_\alpha ({\mbox{\boldmath$k$}})$ in the system with $N=32$
for several choices of $U/t$ 
along the symmetry lines in the B.z.
}
\label{fig6}
  \end{center}
\end{figure}

We now discuss the instability of the insulating state against conventional spin or charge ordered states. To this end, we examine the spin and charge correlations for the system with the largest cluster of $N=32$, where the first-order transition occurs near $U_c\ (N\rightarrow\infty)$ in the thermodynamic limit. We first calculate the momentum distribution function defined by 
\begin{equation}
n_\alpha ({\mbox{\boldmath$k$}})
=\left\{
\begin{array}{ll}
\displaystyle
\frac{1}{2N}\sum_{\sigma}\langle \hat{a}^{\dagger}_{k\alpha\sigma}
\hat{a}_{k\alpha\sigma}\rangle\ &{\rm for}\ {\mbox{\boldmath$k$}}\neq(\pi,\pi), \\
\displaystyle
\frac{1}{4N}\sum_{\sigma,\beta}\langle \hat{a}^{\dagger}_{k\beta\sigma}
\hat{a}_{k\beta\sigma}\rangle\ &{\rm for}\ {\mbox{\boldmath$k$}}=(\pi,\pi).
\end{array}
\right.
\end{equation}
The obtained results are shown in Fig. \ref{fig6}. 
Since the noninteracting system has the quasi-Fermi point at $(\pi,\pi)$, a discontinuity should appear in the momentum distribution for the dispersive band $n_2({\mbox{\boldmath$k$}})$. We indeed observe it for $U<U_c$, while this singularity disappears for $U>U_c$, in accordance with the fact that the MI transition occurs at $U=U_c$, as discussed above.

\begin{figure}[htb]
  \begin{center}
\includegraphics[width=0.45\textwidth]{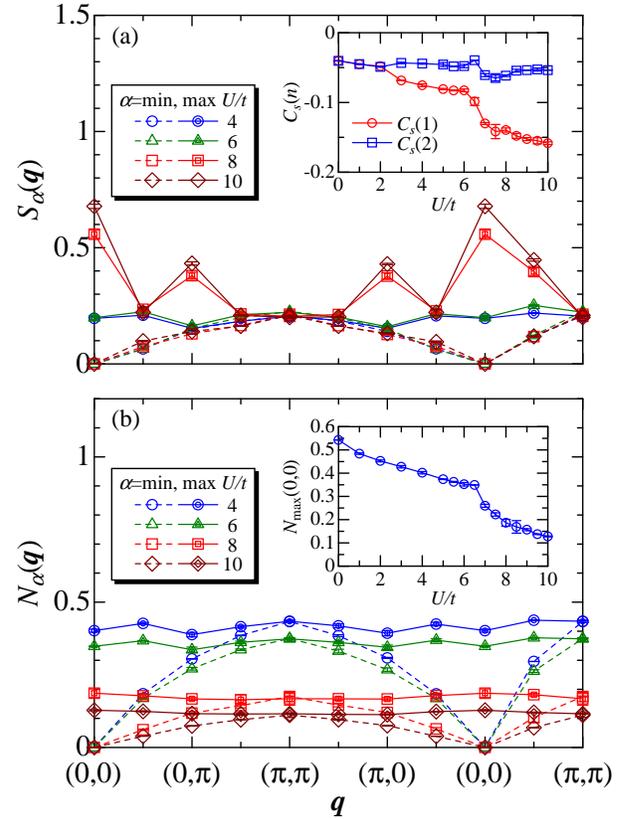}
\caption{(Color online) 
(a) [(b)] Equal-time correlation function 
$S_{\rm \alpha}({\mbox{\boldmath$q$}})$ 
[$N_{\rm \alpha}({\mbox{\boldmath$q$}})$] 
in the system with $N=32$ for different choices of $U/t$ 
along the symmetry lines in the B.z. 
The inset of (a) shows spin correlation $C_s(n)$. 
$N_{\rm max}(0,0)$ as a function of $U/t$ is shown in the inset of (b). 
We note that $4 S_{\alpha}({\mbox{\boldmath$q$}})=
N_{\alpha}({\mbox{\boldmath$q$}})$ at $U/t=0$.
}
\label{fig7}
  \end{center}
\end{figure}

We next consider the equal-time correlation functions in spin and charge sectors, which are given by  
\begin{eqnarray}
S_{mm^{'}}
({\mbox{\boldmath$q$}})
&=&\frac{2}{3N}
\sum_{i,j=1}^{N/2}
\langle 
\hat{\mbox{\boldmath$S$}}_{im}\cdot 
\hat{\mbox{\boldmath$S$}}_{jm'}
\rangle\nonumber\\ 
&\times& 
e^{i{\mbox{\boldmath$q$}\cdot 
(\mbox{\boldmath$R$}_{im}-\mbox{\boldmath$R$}_{jm^{'}})}},\\
N_{mm^{'}}
({\mbox{\boldmath$q$}})
&=&\frac{2}{N}
\sum_{i,j=1}^{N/2}
\left(
\langle \hat{n}_{im}\hat{n}_{jm^{'}}\rangle-\langle \hat{n}_{im}\rangle\langle \hat{n}_{jm^{'}}\rangle
\right)\nonumber\\ 
&\times& e^{i{\mbox{\boldmath$q$}}\cdot 
(\mbox{\boldmath$R$}_{im}-\mbox{\boldmath$R$}_{jm^{'}})},
\end{eqnarray}
where $\hat{n}_{im}=\hat{n}_{im\uparrow}+\hat{n}_{im\downarrow}$ and $\mbox{\boldmath$R$}_{im}$ represents the position of the  $i$-th unit cell in the $m$-th sublattice. Diagonalizing the $2\times2$ matrix, we obtain $S_{\alpha}({\mbox{\boldmath$q$}})$ and $N_{\alpha}({\mbox{\boldmath$q$}})$ $(\alpha={\rm max,\ min})$, as shown in Fig. \ref{fig7}. It is found that spin and charge correlations are little changed in the case $U<U_c$. On the other hand, further increase in the interaction $U$ leads to totally different behavior in the spin correlation function. It is found that when $U>U_c$, $S_{max}({\mbox{\boldmath$q$}})$ is enhanced at ${\mbox{\boldmath$q$}}=(0,0)$ although it never diverges even in the thermodynamic limit. This may suggest that short-range spin correlations are enhanced in the insulating phase. To clarify this point, we also calculate the site-dependent spin correlation function defined by
\begin{eqnarray}
C_s(n)
&=&\frac{1}{N}\frac{1}{N_n}
\sum_{i=1}^{N}\sum_{\tau_n=1}^{N_n}
\langle 
\hat{\mbox{\boldmath$S$}}_{i}\cdot 
\hat{\mbox{\boldmath$S$}}_{i+\tau_n}\rangle,
\end{eqnarray}
where $\tau_n$ labels the $n$-th neighbor site connected by transfer integral $t$ and $N_n$ is the number of them ($N_1=4$, $N_2=2$). We find that both $C_s(1)$ and $C_s(2)$ are always negative as shown in the inset of Fig. \ref{fig7} (a).
Note that the spin correlations $C_s(2)$ are negative even in the strong coupling limit, implying that antiferromagnetic correlations are suppressed due to geometrical frustration.
As for the charge sector, the increase in $U/t$ gradually suppresses charge correlations, as shown in Fig. \ref{fig7} (b). In addition, we find the monotonic decrease in $N_{\rm max}(0,0)$ in the inset. Therefore, there is no tendency to stabilize the charge ordered state with ${\mbox{\boldmath$q$}}=(0,0)$.\cite{Fujimoto}

\begin{figure}[htb]
  \begin{center}
\includegraphics[width=0.45\textwidth]{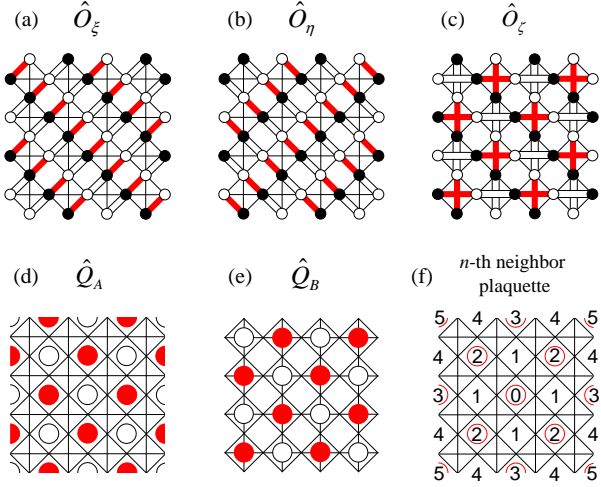}
\caption{(Color online) 
(a)-(c) represent the dimer ordering patterns. 
Thick (double) lines represent positive (negative) signs
 for the dimer correlation function eq. (\ref{D}).
(d) and (e) represent the plaquette ordering patterns. 
Filled (open) circles represent positive (negative) signs for the plaquette
correlation function eq. (\ref{P}).
(f) represents the $n$-th neighbor plaquettes, where the reference plaquette is labeled by zero.}
\label{fig8}
  \end{center}
\end{figure}

\begin{figure}[htb]
  \begin{center}
\includegraphics[width=0.45\textwidth]{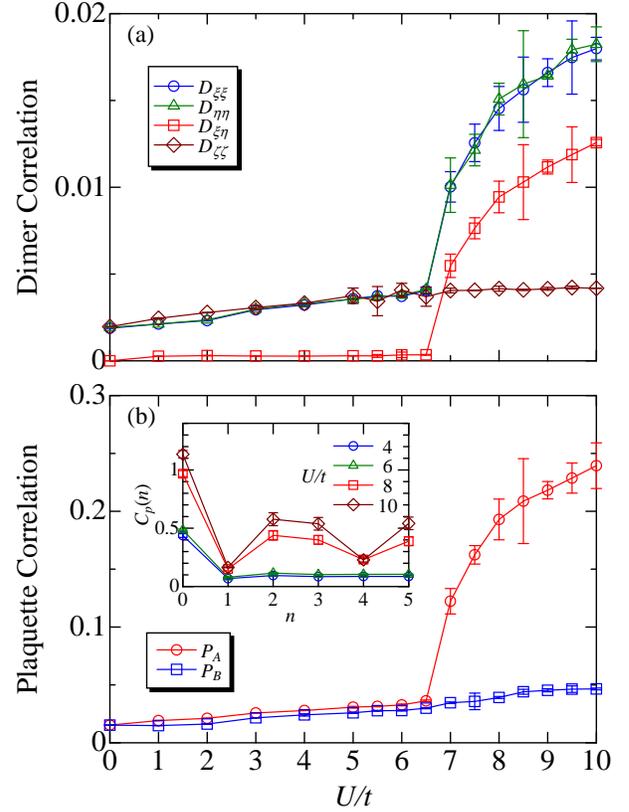}
\caption{(Color online) (a) The dimer correlation functions $D_{\alpha\beta}$ and (b) the plaquette correlation functions $P_{A(B)}$ on the $N=32$ lattice as a function of $U/t$. 
The inset in (b) is the $n$-th neighbor plaquette correlation $C_p(n)$.}
\label{fig9}
  \end{center}
\end{figure}

To discuss the spin configuration of the insulating phase in detail, we calculate the dimer correlation function $D_{\alpha\beta}$ and the plaquette correlation function $P_{A(B)}$ defined by
\begin{eqnarray}
D_{\alpha\beta}&=&\langle \hat{O}_{\alpha}\hat{O}_{\beta}\rangle,\label{D}\\
\hat{O}_{\alpha}&=&\frac{1}{N}\sum_{\stackrel{i=1}{{\rm pattern}\ \alpha}}^{N}(-1)^i\hat{d}_i,\ \ {\rm for}\ \ \alpha=\xi,\ \eta,\ \zeta,
\end{eqnarray}
and
\begin{eqnarray}
P_{A(B)}
&=&\langle \hat{Q}_{A(B)}^2\rangle, \label{P}\\
\hat{Q}_{A(B)}&=&\frac{2}{N}
\sum_{\stackrel{i=1}{{\rm pattern}\ A(B)}}^{N/2}(-1)^i\hat{p}_i,
\end{eqnarray}
where $\hat{d}_i(=\hat{\mbox{\boldmath$S$}}_{i}\cdot 
\hat{\mbox{\boldmath$S$}}_{i+{\delta}})$ is the $i$-th dimer operator and
$\hat{p}_i[=
(\hat{\mbox{\boldmath$S$}}_{\alpha_i}+
\hat{\mbox{\boldmath$S$}}_{\gamma_i})\cdot (
\hat{\mbox{\boldmath$S$}}_{\beta_i}+
\hat{\mbox{\boldmath$S$}}_{\delta_i})]$ is the $i$th plaquette operator.
The patterns for dimers and plaquettes, and their signs $(-1)^i$ are schematically shown in Figs. \ref{fig8} (a)-(e). 

Figure \ref{fig9} (a) shows the $U/t$ dependence of the dimer correlations for several possible configurations. We find that the dimer correlations are negligibly small for all the patterns when $U<U_c$. On the other hand, beyond $U=U_c$, the dimer correlations for some patterns are suddenly enhanced, while $D_{\zeta\zeta}$
 remains small. Namely, a crossed-dimer valence-bond ordered state [see Fig. \ref{fig8} (c)], which is realized in the weakly coupled Heisenberg chains on the checkerboard lattice \cite{cvbc}, is not stabilized in the strong coupling limit. The existence of three dimer correlations concludes that the ordinary dimer ordered state is not stabilized but the plaquette ordered state with pattern $A$ emerges instead. It is indeed seen that the positive $D_{\xi\eta}$ appears in the plaquette ordered state with $A$ configuration.

To confirm the above results, we also calculate the plaquette-singlet correlations, as shown in Fig. \ref{fig9} (b). It is seen that the plaquette correlations with pattern $A$ are enhanced in the region $U>U_c$. Therefore, the quantum phase transition breaks the translational symmetry, leading to the formation of a plaquette-singlet ordered state. In fact, the plaquette order parameter alternates spatially in the insulating phase ($U>U_c$), as shown in the inset of Fig. \ref{fig9}. Here, the $n$-th neighbor plaquette correlations are defined by
\begin{equation}
C_p(n)=\frac{2}{N}\frac{1}{N_n}
\sum_{i=1}^{N/2}\sum_{\tau_n=1}^{N_n}
\langle\hat{p}_i\hat{p}_{i+\tau_n}\rangle,
\end{equation}
where $\tau_n$ runs the $n$-th neighbor plaquettes and $N_n$ is the number of them (Fig. \ref{fig8} (f)). These results are consistent with those in the Heisenberg limit\cite{Fouet}, where the plaquette valence-bond crystal phase is stabilized. We thus end up with the conclusion that in the Hubbard model on the checkerboard lattice, the normal metallic state is realized for small $U$, while the plaquette-ordered insulating state is for large $U$. The first-order transition occurs between these phases at $U_c/t=6.75\pm0.25$. 

Here, we wish to make a brief comment on our extrapolation scheme to the thermodynamic limit. In this paper, we have carried out the extrapolation in a scaling form of $N^{-3/2}$ for the physical quantities, which yields accurate results. In fact, we find in Fig. \ref{fig5} that the obtained results are in good agreement with those obtained by the weak and strong coupling techniques such as the Green's function approach with the self-energy up to second-order in $U$ and the fourth-order plaquette expansion around the configuration shown in Fig. \ref{fig8} (d). We thus confirm that our PIRG method with a new iteration scheme works well to obtain the reliable results for the Hubbard model on the checkerboard lattice.

\section{Summary}

We have studied the Hubbard model on the checkerboard lattice at half filling by means of the PIRG method. When the method has been naively applied to our frustrated model, we have encountered a serious problem that the approximate ground state is sometimes trapped in local minima, leading to unreliable results. To overcome this, we have proposed a new iteration scheme in the PIRG method, and have demonstrated that the satisfactory convergence is achieved in much smaller iteration steps. We believe that our iteration scheme can be generally applied to other frustrated electron systems for accelerating the convergence and improving the numerical accuracy. 

It has been found that the increase in the Hubbard interaction yields the first-order metal-insulator transition at $U_c/t=6.75\pm0.25$, where the jump singularity appears in the double occupancy. Furthermore, we have calculated the dimer and plaquette correlation functions to clarify that the system is driven to the plaquette ordered state.

We have focused on the isotropic checkerboard lattice model in this paper. If the magnitude of the diagonal hopping is varied, magnetic correlations are enhanced, which should yield a rich phase diagram. This issue on the anisotropic checkerboard lattice is important to study the role of geometrical frustration systematically, which is now under consideration.
It also remains interesting to investigate finite-temperature Mott transitions, which should provide further invaluable information about frustrated electrons on the checkerboard lattice.

\section*{Acknowledgment}

The authors thank Y. Imai, and T. Ohashi for valuable discussions. 
Parts of computations were done at the Supercomputer Center at the
Institute for Solid State Physics, University of Tokyo. 
The work is partly supported by a Grant-in-Aid from the Ministry of Education, 
Culture, Sports, Science, and Technology of Japan 
[20740194 (A.K.),  20029013 (N.K.) and 19014013 (N.K.)]. 
T.Y. is supported by the Japan Society for the Promotion of Science.


\begin{thebibliography}{99} 
\bibitem{exp3}
S. Kondo, D. C. Johnston, C. A. Swenson, F. Borsa, A. V. Mahajan, L. L. Miller, T. Gu, A. I. Goldman, M. B. Maple, D. A. Gajewski, E. J. Freeman, N. R. Dilley, R. P. Dickey, J. Merrin, K. Kojima, G. M. Luke, Y. J. Uemura, O. Chmaissem, and J. D. Jorgensen: Phys. Rev. Lett. {\bf 78} (1997) 3729 .

\bibitem{exp1} T. Takeda, M. Nagata, H. Kobayashi, R. Kanno, Y. Kawamoto, M. Takano, T. Kamiyama, F. Izumi and A. W. Sleight: J. Solid State Chem. 
{\bf 140} (1998) 182.

\bibitem{exp2} H. Sakai, M. Kato, K. Yoshimura and K. Kosuge: J. Phys. Soc. Jpn. 
{\bf 71} (2002) 422.

\bibitem{Isoda}
M. Isoda and S. Mori: J. Phys. Soc. Jpn. {\bf 69} (2000) 1509. 

\bibitem{Fujimoto2}
S. Fujimoto: Phys. Rev. B {\bf 64} (2001) 085102.

\bibitem{Fujimoto}
S. Fujimoto: Phys. Rev. Lett. {\bf 89} (2002) 226402.

\bibitem{tJ}
D. Poilblanc: Phys. Rev. Lett. {\bf 93} (2004) 197204.

\bibitem{LiV1}
C. Lacroix: Can. J. Phys. {\bf 79} (2001) 1469.

\bibitem{LiV2}
N. Shannon: Eur. Phys. J. B {\bf 27} (2001) 527.

\bibitem{LiV3}
P. Fulde, A. N. Yaresko, A. A. Zvyagin and Y. Grin: Europhys. Lett. {\bf 54} (2001) 779.

\bibitem{LiV4}
S. Burdin, D. R. Grempel and A. Georges: Phys. Rev. B {\bf 66} (2002) 045111.

\bibitem{LiV5}
J. Hopkinson and P. Coleman: Phys. Rev. Lett. {\bf 89} (2002) 267201.

\bibitem{LiV6}
S. Fujimoto: Phys. Rev. B {\bf 65} (2002) 155108.

\bibitem{LiV7}
H. Tsunetsugu: J. Phys. Soc. Jpn {\bf 71} (2002) 1844.

\bibitem{LiV8}
Y. Yamashita and K. Ueda: Phys. Rev. B {\bf 67} (2003) 195107.

\bibitem{LiV9}
M. S. Laad, L. Craco and E. M\"uller-Hartmann: Phys. Rev. B {\bf 67} (2003) 033105.

\bibitem{label5}
T. Yoshioka, A. Koga and N. Kawakami: Physica B {\bf 378} (2006) 294.

\bibitem{label6}
A. Koga, T. Yoshioka, N. Kawakami, and H. Yokoyama: Physica C {\bf 460} (2007) 1070.

\bibitem{Fouet}
J.-B. Fouet, M. Mambrini, P. Sindzingre and C. Lhuillier: Phys. Rev. B {\bf 67} (2003) 054411.

\bibitem{graph}
A. Mielke: J. Phys. A {\bf 24} (1991) L73.

\bibitem{PIRG0}
M. Imada and T. Kashima: J. Phys. Soc. Jpn. {\bf 69} (2000) 2723.

\bibitem{PIRG1}
T. Kashima and M. Imada: J. Phys. Soc. Jpn. {\bf 70} (2001) 2287.

\bibitem{QMCstart}
N. Furukawa and M. Imada: J. Phys. Soc. Jpn. {\bf 60} (1991) 3669.

\bibitem{Sorella} 
S. Sorella: Phys. Rev. B {\bf 64} (2001) 024512.

\bibitem{PIRG2} 
T. Mizusaki and M. Imada: Phys. Rev. B {\bf 69} (2004) 125110.

\bibitem{PIRG7} 
T. Mizusaki and M. Imada: Phys. Rev. B {\bf 74} (2006) 014421.

\bibitem{cvbc}
O. A. Starykh, A. Furusaki and L. Balents: Phys. Rev. B {\bf 72} (2005) 094416

\end{thebibliography}
\end{document}